\documentclass[a4paper,8pt,onecolumn]{article}%
\usepackage{amsmath}
\usepackage{amssymb}
\usepackage{amsthm}
\usepackage{graphicx}
\usepackage{subfigure}
\usepackage[latin1]{inputenc}
\usepackage{color}%
\usepackage{amsfonts}
\providecommand{\U}[1]{\protect\rule{.1in}{.1in}}

\begin{document}
\title{Stabilization of a linear nanomechanical oscillator to its ultimate thermodynamic limit}

\author{{\normalsize E. Gavartin$^{\mathrm{1}}$, P. Verlot$^{\mathrm{1}}\footnote{Corresponding author; Email: pierre.verlot@epfl.ch}\,$  and T.J.
Kippenberg$^{\mathrm{1,2}}\footnote{Corresponding author; Email: tobias.kippenberg@epfl.ch}$}\\{\normalsize $^{\mathrm{1}}$Ecole Polytechnique F{\'e}d{\'e}rale de Lausanne,
EPFL, 1015 Lausanne, Switzerland}\\{\normalsize $^{\mathrm{2}}$Max-Planck-Institut f{\"u}r Quantenoptik,
Hans-Kopfermann-Stra{\ss}e 1, 85748 Garching, Germany}}

\maketitle

\begin{abstract}
The rapid development of micro- and nanooscillators in the past decade has led to the emergence of novel sensors that are opening new frontiers in both applied and fundamental science. The potential of these novel devices is, however, strongly limited by their increased sensitivity to external perturbations. We report a non-invasive optomechanical nano-stabilization technique and apply the method to stabilize a linear nanomechanical beam at its ultimate thermodynamic limit at room temperature. The reported ability to stabilize a mechanical oscillator to the thermodynamic limit can be extended to a variety of systems and increases the sensitivity range of nanosensors in both fundamental and applied studies.
\end{abstract}

Thermal noise is well known to be a ubiquitous and fundamental limit for the sensitivity of micro- and nanomechanical oscillators  \cite{Albrecht1991,Cleland2002,Ekinci2005}. This noise arises from the unavoidable coupling of the mechanical device to a thermal bath \cite{saulson1990thermal}, which is responsible for an incoherent motion with random amplitude and phase \cite{Briant2003}. A number of precision measurements based on using nanomechanical resonators, such as gradient force \cite{giessibl1995atomic,Rugar2004}, mass \cite{Chaste2012,Jensen2008} and charge \cite{Cleland1998} detection, or time and frequency control \cite{Nguyen2007}, consist in detecting linear changes induced by the measured system on the phase of the coherently driven nanomechanical sensor \cite{Albrecht1991,binnig1986atomic}. These applications are thus theoretically limited by the random phase fluctuations related to thermal motion. Decreasing the effect of thermal noise can be accomplished by increasing the relative contribution of the coherent drive \cite{Albrecht1991}. However, the latter has to be kept below a certain threshold, above which the mechanical oscillator becomes nonlinear \cite{Ekinci2005,Nayfeh1979}: At this threshold, thermal noise constitutes an irreducible sensing limit for linear nanomechanical measurements. In practice, even when driven close to their nonlinear onset, nanomechanical resonators presently operate far away from the thermal limit \cite{Chaste2012,Feng2008,Lee2010,Fong2012}. This is due to a plethora of external perturbation sources acting on the oscillator \cite{Cleland2002}, such as temperature fluctuations \cite{ekinci2004ultimate}, adsorption-desorption noise \cite{Yang2011} or molecular diffusion along the oscillator \cite{Atalya2011}. While several proposals have been made \cite{Yurke1995,Kenig2012} and implemented \cite{Lopez2012,villanueva2011nanoscale} to decrease excess phase noise, they are based on oscillators being driven into the nonlinear regime and require specific operating conditions, thus making the schemes difficult to adapt for a general class of resonators. In contrast, here we introduce an approach that generally applies to linear nanomechanical resonators. This approach employs an auxiliary mechanical mode as a frequency noise detector, whose output is used to stabilize the frequency of the fundamental out-of-plane mode of a nanomechanical beam. Its experimental implementation enables quasi-optimal linear operation at room temperature, with an almost perfect external phase noise cancellation, as demonstrated for the nanomechanical beam driven just below its nonlinear onset. Hence, our system operates close to its ultimate thermodynamic linear sensing limit. Our scheme can be readily incorporated into a variety of other systems used for applications in \cite{Rugar2004,Chaste2012,Jensen2008}, frequency control \cite{Nguyen2007} or fundamental studies \cite{verlot2009scheme,Arcizet2011}.

Our system consists of an integrated hybrid optomechanical transducer described elsewhere \cite{Gavartin2012}. Briefly, it consists of a $90\,\mu \mathrm{m}\times700\,\mathrm{nm}\times100\,\mathrm{nm}$ low dissipation nanomechanical beam made of high-stress silicon nitride (Si$_{3}$N$_{4}$) placed in the near-field of a high-Q whispering gallery mode (WGM) confined in a disk-shaped microcavity (cf. Fig. 1(a)). Laser light is coupled in and out of the oscillator with a tapered fiber. Mechanical motion is read out using the optomechanical interaction by placing the laser frequency at mid-transmission of the optical mode, such that the phase fluctuations of light induced by the mechanical motion are directly transformed into amplitude fluctuations readily detected with a photodiode \cite{Cai2000}.

Our stabilization scheme is based on the intuition that most of the observed external frequency noise in our system arises from perturbations affecting the entire beam geometry (e.g. temperature fluctuations causing causing changes in length, strain among others). Under such conditions, the induced frequency fluctuations of the various eigenmodes are expected to be highly correlated. One of these modes can thus be used as a "noise detector" whose output serves as an error signal in a feedback mechanism applied to the beam, enabling a cancellation of the geometry variations and thereby a stabilization of all eigenmodes (cf. Fig. 2(a)). In the following our analysis will focus more specifically on two modes, which are the lowest-order out-of-plane mode, which we seek to stabilize, and the lowest-order in-plane mode, which is used as the noise detector. These two modes (referred to as mode $1$ and $2$, respectively, in the following) have resonance frequencies $\nu_{\mathrm{M},1}=\Omega_{\mathrm{M},1}/2\pi=2.84\,\mathrm{MHz}$ and $\nu_{\mathrm{M},2}=\Omega_{\mathrm{M},2}/2\pi=3.105\,\mathrm{MHz}$ as well as linewidths  of $\Gamma_{\mathrm{M},1}/2\pi=7.5\,\mathrm{Hz}$ and $\Gamma_{\mathrm{M},2}/2\pi=48\,\mathrm{Hz}$. Note that using the in-plane mode as the noise detector has several advantages: First, the accurate position resolution of the beam's intrinsic motion enables to achieve a large bandwidth ($\nu_{\mathrm{D}}\simeq 20\times \Gamma_{\rm{M},2}/2\pi\simeq1\,\mathrm{kHz}$, cf. Fig. 1(d)) over which frequency fluctuations can be detected with thermal limited imprecision. Second, this geometry enables a selective exposure of the perpendicular out-of-plane mode to external signals, thus preserving its sensing capabilities when feedback is turned on. Experimentally, the frequency fluctuations of each mode are detected via its out-of-phase response to an external radiation pressure coherent drive with a frequency of $\Omega_{\mathrm{F},i}\,(i\in\{1,2\})$, similar to a phase-locked-loop detection. We choose the driving strengths such that both modes have an amplitude of just below the respective nonlinear thresholds. Both measured out-of-phase quadratures for mode 1 and 2 are shown in Figure 2(d) with the demodulated response plotted versus the detuning $\omega_i/2\pi=(\Omega_{\mathrm{F},i}-\Omega_{\mathrm{M},i})/2\pi$ between the driving and the resonance frequency. By putting the external drive close to the resonance frequency ($\Omega_{\mathrm{F},i}\simeq\Omega_{\mathrm{M},i}$), we can monitor the resonance frequency as small changes on the order of $\delta\Omega_i \lesssim\Gamma_{\rm{M},i}$ that are proportional to the out-of-phase quadrature signals. Our feedback mechanism relies on DC-photothermal actuation applied to the beam by sending a third amplitude modulated laser tone (control beam) with input power in the $\mathrm{mW}$ range (see Fig. 1(e)). The frequency shift follows linearly the optical input power as shown in Figure 2(b) for mode 1.

Figure 2(c) shows the frequency response of modes $1$ and $2$ to a triangular wave function applied to the control beam. One can already note the similarities of the fluctuations superimposed to the response triangles in both cases, indicating the common origin of the frequency noise over the various eigenmodes.

These noise correlations can be quantitatively confirmed by determining the frequency correlations $\mathcal{C}_{1-2}[\omega]=\langle\delta\nu_1[\omega]\delta\nu_2^*[\omega]\rangle/\left(S_{\nu_1}[\omega]S_{\nu_2}[\omega]\right)^{1/2}$ (with $\delta\nu_i[\omega]$ denoting the Fourier transform of the frequency fluctuations of mode $i$, $S_{\nu_i}$ its spectrum and $\langle...\rangle$ statistical average), which are represented in Figure 3(c) (inset). It should be emphasized that in absence of external frequency noise, the frequency thermal imprecisions over modes $1$ and $2$ are expected to be uncorrelated: the high level of correlations is therefore a signature of the presence of external frequency noise. Experimentally, this excess of frequency noise is determined by comparing the phase noise spectrum $S_{\phi\phi,i}$ of mode $i$ to its thermal equivalent imprecision, which in the strong drive limit is given by \cite{Cleland2002,Briant2003}

\begin{eqnarray}
S^{\rm{th}}_{\phi\phi,i}\left[\omega\right]&\simeq& \frac{1}{\langle X_{\mathrm{osc},i}^{2}\rangle}\times\frac{\Gamma_{\mathrm{M},i}k_{\rm{B}}T}{m_{\rm{eff},i}\Omega^{2}_{\mathrm{M},i}\left(\omega^{2}+\Gamma_{\mathrm{M},i}^{2}/4\right)},
\label{eq:phasenoise}
\end{eqnarray}

with $\langle X_{\mathrm{osc},i}^2\rangle$ as the variance of the driven amplitude, $m_{\mathrm{eff},i}$ as the effective mass, $k_\mathrm{B}$ as Boltzmann's constant, $T$ as the temperature and $\omega$ as the Fourier frequency. Equation \ref{eq:phasenoise} simply reflects the fact that phase noise can be interpreted as the "angle" under which the thermal distribution appears from the center of the phase space (see Fig. \ref{Fig3}(b)). The time evolution of the frequency drift $\delta\nu_{\rm{M},i}$, as determined from the mechanical out-of-phase response, is shown in Figure 3(a) for mode 1 and mode 2. It is clearly evident that the frequency noise is strongly decreased for the stabilized case. These time traces are used to obtain the phase noise spectrum via a Fourier transform. The results are shown in Figure 3(c), both in the non-stabilized (red) and stabilized (blue) cases. Dark colored curves correspond to mode $1$ and light colored to mode $2$. The results are calibrated using the inferred driven root-mean-square amplitudes $\langle X_{\mathrm{osc},1}^2\rangle^{\frac{1}{2}}=1.7\,\mathrm{nm}$ and $\langle X_{\mathrm{osc},2}^2\rangle^{\frac{1}{2}}=4.1\,\mathrm{nm}$, just below their respective nonlinear thresholds. When the system is not stabilized, the phase noise of both modes is clearly above the thermal limit. With the stabilization scheme turned on, the phase noise of mode 1 is reduced to the thermal limit in the frequency range of $\sim$10-50 Hz. The spectrum of mode 2 implies that its phase noise was reduced to even below the thermal limit, which is clearly unphysical. This artifact arises from squashing, which is generally observed in an in-loop measurement where the feedback gain is so large that the reinjected background -  in our case the thermally induced frequency fluctuations - is the dominating part of the error signal. As the spectrum of mode 1 is obtained in an out-of-loop measurement, reinjection of background would manifest itself correctly as an increase in the phase noise spectrum. Since the thermal frequency noise background $S^{\rm{th}}_{\omega\omega,i}=\omega^{2}S^{\rm{th}}_{\phi\phi,i}$ is smaller for mode 2 (28 $\rm{mHz}/\sqrt{\rm{Hz}}$) in comparison to mode 1 (41 $\rm{mHz}/\sqrt{\rm{Hz}}$), the reinjection of noise is limited for mode 1 - the mode of interest - to a level of 1.5 dB with respect to the thermal limit.

We also observe the effect of stabilization on the correlations between the frequency noise of both modes. As noted above, noise suppression is expected to correspond to a significant decrease of the frequency noise correlations, as the noises resulting from fundamental thermodynamic fluctuations are uncorrelated. We confirm this experimentally as shown in the inset in Figure 3(c).

To access the long-term stability for mode $1$, we determine its Allan deviation. In order to avoid measurement artifacts, the mechanical motion is no longer driven using an external frequency reference, but is maintained in self-oscillation using a feedback circuit \cite{Cohadon1999,Gavartin2012} (see Fig. 1(e)). The readout signal is fed into a frequency counter ({\sc Agilent 53230}), which determines the Allan deviation as $\langle\delta \nu_{\rm{M},1}/\nu_{\rm{M},1}\rangle _{\tau}=\left[\frac{1}{2 (N-1)}\sum_{k=1}^{N}\left(\left(\bar{\nu}_{\mathrm{M},k+1}-\bar{\nu}_{\mathrm{M},k}\right)/\nu_{M}\right)^{2}\right]^{1/2}$, with $\bar{\nu}_{\mathrm{M},k}$ as the averaged frequency in the $k$th discrete time interval over the gate time $\tau$.

The obtained results are shown in Figure 3(d), where the red and blue dots represent the values of the fractional Allan deviation of mode $1$ as a function of gate time in the non-stabilized and stabilized cases, respectively. The slow fluctuations are significantly reduced in presence of stabilization, with the noise suppression even exceeding $20\,\mathrm{dB}$ at the second scale. The achieved long-term stability is further compared to the theoretical limits of our system. For a self-sustained mechanical resonator, the thermo-mechanically limited Allan variance $\sigma_1(\tau)=\langle\delta\nu_{\mathrm{M},1}^{\mathrm{th}}/\nu_{\mathrm{M},1}\rangle_{\tau}$ can be shown to be given by (see Supplementary section S2.2):

\begin{eqnarray}
\sigma_1(\tau)&=&\left(\frac{\langle X_{\mathrm{th},1}^2\rangle}{\langle X_{\mathrm{osc},1}^2\rangle}\times \frac{1}{Q_{\mathrm{M},1}\Omega_{\mathrm{M},1}\tau}\right)^{1/2},\label{eq:stab-allan}
\end{eqnarray}

The data of Figure 3(d) were acquired in presence of a self-sustained displacement $\langle X_{\mathrm{osc},1}^2\rangle^{1/2}\simeq100\langle X_{\mathrm{th},1}^2\rangle^{1/2}$.

This analysis shows that our stabilization scheme enables reducing external frequency noise close to the thermal limit, but not closer than a factor of $2$ (that is $6\,\mathrm{dB}$ in power), which disagrees with the spectral data of Figure 3(c). This can be explained by a reduced sensitivity of noise detection: Mode $2$ being externally driven, the expression of its thermal limited Allan deviation differs from mode $1$ and is described by (see Supplementary section 2.2):
\begin{eqnarray}
\sigma_2(\tau)&=&\left(
\frac{\langle X_{\mathrm{th},2}^2\rangle}{\langle X_{\mathrm{osc},2}^2\rangle}\times \frac{3+e^{-\Gamma_{\mathrm{M},2}\tau}-4e^{-\frac{\Gamma_{\mathrm{M},2}\tau}{2}}}{(\Omega_{\mathrm{M},2}\tau)^2}\right)^{\frac{1}{2}}.\nonumber\\
\label{eq:stab-allan-2}
\end{eqnarray}
With $\langle X_{\mathrm{osc},2}^2\rangle^{1/2}\simeq10^2\langle X_{\mathrm{th},2}^2\rangle^{1/2}$, the noise detection imprecision $\langle \delta\nu_{\mathrm{M},2}^{\mathrm{th}}\rangle_{\tau\simeq0.1\,\mathrm{s}}\simeq10^{-8}\nu_{\mathrm{M},2}$ is comparable to $\langle \delta\nu_{\mathrm{M},1}^{\mathrm{th}}\rangle_{\tau\simeq0.1\,\mathrm{s}}\simeq10^{-8}\nu_{\mathrm{M},1}$ (Eq. \ref{eq:stab-allan}), and therefore has to be taken into account when determining the stabilization limit, which is given by:

\begin{eqnarray}
\langle \delta\nu_{\mathrm{M},1}^{\mathrm{lim}}\rangle_{\tau}&=&\left(\langle \delta\nu_{\mathrm{M},1}^{\mathrm{th}}\rangle_{\tau}^2+\langle \delta\nu_{\mathrm{M},2}^{\mathrm{th}}\rangle_{\tau}^2\right)^{1/2}.\label{eq:lim-total}
\end{eqnarray}

The thermal limit for mode $2$ $\langle\delta\nu_{\mathrm{M},2}^{\mathrm{th}}\rangle_{\tau}$ as well as the overall stabilization limit given by Eq. \ref{eq:lim-total} are represented in Figure 3(d), showing very good agreement with the experimental data up to gate times in the $100\,\mathrm{ms}$ range. On longer timescales, the frequency fluctuations are stabilized to a flat level in the $10\,\mathrm{ppb}$ range, which we attribute to insufficient gain.

Finally, we analyze the frequency stability and its improvement with the stabilization scheme by using a third independent method based on a recent proposal \cite{Maizelis2011}. Briefly, this method is based on the simultaneous measurement of both in-phase and out-of-phase quadratures $X_1(t)$ and $X_2(t)$ of the coherently driven mechanical motion. These quadratures are used to reconstruct the complex motion trajectory $u(t)=X_1(t)+iX_2(t)$ whose moments $\langle u^n\rangle$ can be shown to be insensitive to thermal noise, enabling the extraction and quantitative study of external frequency noise. Thus, the presence of external frequency fluctuations $\delta\Omega_{\mathrm{M}}(t)$ is revealed by an anomaly of the ratio $r=\langle u^2\rangle/\langle u\rangle^2$, which falls below unity by an amount proportional to the frequency noise power. Indeed, the quadratures of the coherently driven motion write as the sum of a steady and a fluctuating part $X_{1,2}(t)=\langle X_{1,2}\rangle+\delta X_{1,2}(t)$. A straight forward expansion gives $\langle u^2\rangle=\langle u\rangle^2+\langle(\delta X_1)^2\rangle-\langle(\delta X_2)^2\rangle-2i\langle\delta X_1\delta X_2\rangle$. With the thermal contribution to the quadrature fluctuations $\delta X_{1,2}^{\mathrm{th}}$ being uncorrelated and with identical variances, the previous expression becomes $\langle u^2\rangle=\langle u\rangle^2+\langle(\delta X_1^{\mathrm{ext}})^2\rangle-\langle(\delta X_2^{\mathrm{ext}})^2\rangle-2i\langle\delta X_1^{\mathrm{ext}}\delta X_2^{\mathrm{ext}}\rangle$, with $X_{1,2}^{\mathrm{ext}}$ denoting the quadratures fluctuations due to the presence of external frequency noise. For a resonantly driven system, the in-phase quadrature $X_1$ is maximum, and $\delta X_1^{\mathrm{ext}}=\left(\partial X_1/\partial\omega\right)\delta\Omega_{\mathrm{M}}(t)=0$, $\langle u^2\rangle$ is thus reduced to the quantity $\langle\left(\delta X_2^{\mathrm{ext}}\right)^2\rangle\simeq\left(\partial X_2/\partial\omega\right)^2\langle\delta\Omega_{\mathrm{M}}^2\rangle\simeq\frac{4\langle\delta\Omega_{\mathrm{M}}^2\rangle}{\Gamma_{\mathrm{M}}^2}\langle u\rangle^2$. At the mechanical resonance frequency, $r$ gives a direct access to the frequency noise power, with no calibration being required. In a more sophisticated approach \cite{Maizelis2011}, it can be shown that for a weak Gaussian frequency noise, the ratio $r[\omega]$ varies along the driving frequency as

\begin{eqnarray}
r[\omega]&=&1-\frac{S_{\omega\omega}}{\Gamma_{\mathrm{M}}+2i\omega},\label{eq:dykmanrig}
\end{eqnarray}

with $S_{\omega\omega}$ denoting the frequency noise spectral density.
Figure 4 shows the experimental determination of $r[\omega]$ we performed with our system, both in the non-stabilized (Fig. 4(c)) and in the stabilized cases (Fig. 4(c,d)). Figures 4(a) and 4(b) show the realizations of the phase-space trajectory used to determined $r[\omega\simeq0]$ in the non-stabilized and stabilized configurations respectively.

As expected, without stabilization we obtain a discernible dip around the mechanical resonance frequency as shown in Figure 4(c). The dip depth is about $1-r[\omega=0]\simeq4\times10^{-2}$ with the deviations from the expected shape resulting from long-term frequency drifts that change $\delta\omega$. Following Eq. \ref{eq:dykmanrig} such a dip is expected to correspond to a frequency noise $S_{\nu\nu}=S_{\omega\omega}/(2\pi)^2=(218\,\mathrm{mHz}/\sqrt{\mathrm{Hz}})^2$, that is about $15\,\mathrm{dB}$ above the thermal limit $S_{\nu\nu}^{\mathrm{th}}[\omega]=\frac{\omega^2}{4\pi^2}S_{\mathrm{\phi\phi},1}[\omega]\simeq(41\,\mathrm{mHz}/\sqrt{\mathrm{Hz}})^2$, in very good agreement with the results presented in Fig. \ref{Fig3}(c). When the stabilization is switched on, the dip is greatly reduced to a level of $1-r[\omega=0]\simeq6\times10^{-4}$, corresponding to a residual noise $S_{\nu\nu}=(27\,\mathrm{mHz}/\sqrt{\mathrm{Hz}})^2$ that is $1.5\,\mathrm{dB}$ above the expected thermal limit. This value confirms the origin of this remaining imprecision as resulting from squashing, which is expected to generate the very same noise excess as explained above.

In conclusion, we have proposed and implemented a scheme allowing a non-invasive stabilization of a mechanical mode. This technique proves to be remarkably efficient, as shown by the almost perfect noise cancellation for the lowest order out-of-plane mode of a high-Q nanomechanical resonator operated just below its nonlinear threshold, which places our system close to the optimum noise limit at room temperature. Our method is very general and can be readily applied to a number of ultra-sensitive nanomechanical systems: First, our scheme allows using any mechanical mode for detection of frequency noise and we successfully demonstrated noise cancellation using the third-order out-of-plane mode. To apply an error signal one needs a DC restoring force, which is widely available either via photothermal forces in optomechanical systems such as in our work or as electrostatic forces in electrical systems \cite{kozinsky2006tuning,Faust2012,Poggio2008}. Importantly, our scheme fully preserves the detection capabilities of the stabilized system, which we have experimentally verified by showing efficient radiation-pressure induced thermomechanical squeezing \cite{Rugar1991} of the fundamental out-of-plane mode \cite{Gavartin2012b} that was impossible to observe without stabilization due to long-term frequency drifts.

The fabrication was carried out at the Center of MicroNanotechnology (CMi) at EPFL. Funding for this work was provided by the NCCR Quantum Photonics, the SNF, DARPA Orchid and an ERC.\newline

\clearpage
\onecolumn

\begin{figure}[ptb]
\includegraphics[scale=.12]{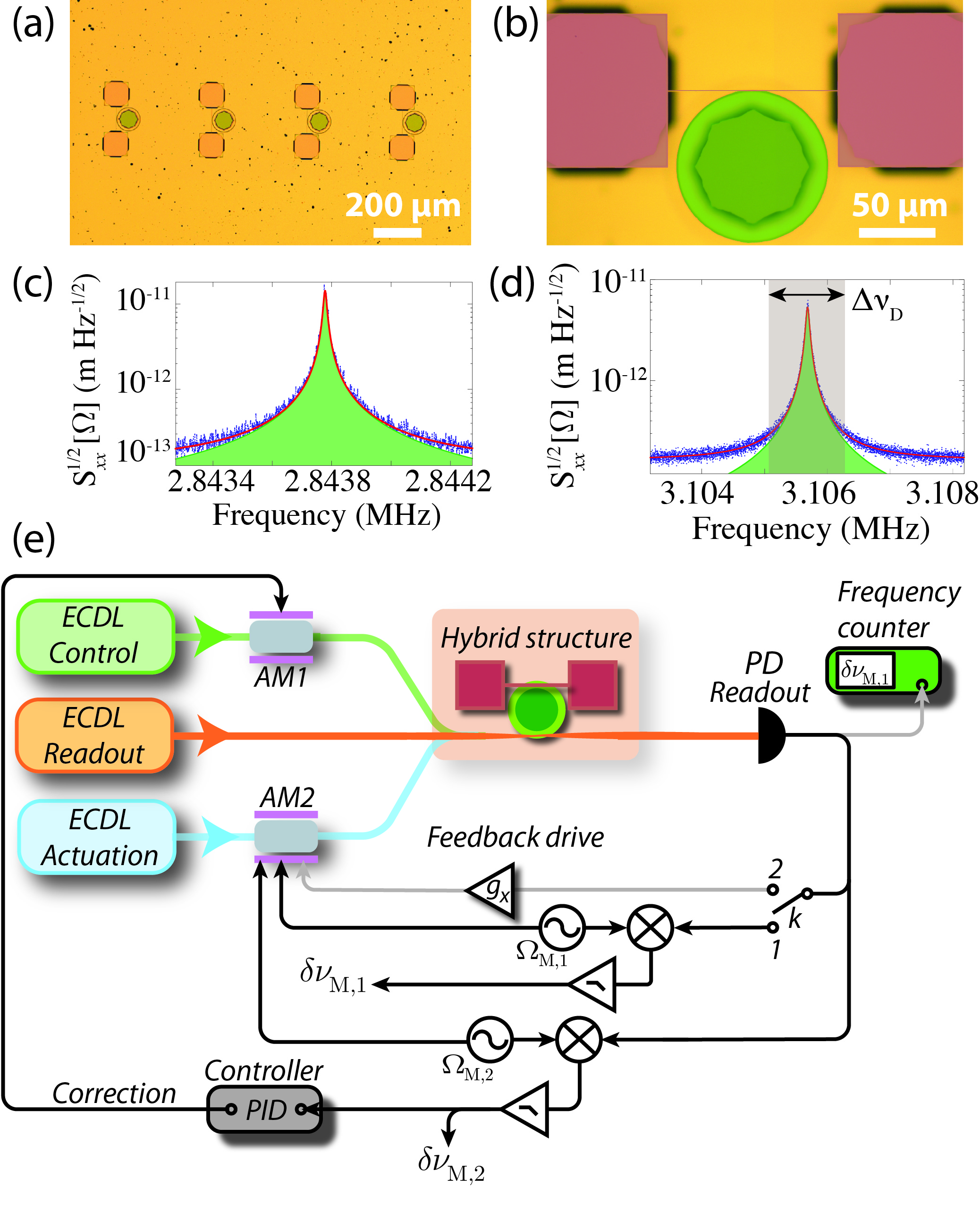}
\caption{\textbf{The hybrid optonanomechanical system and experimental setup.} (a) Optical micrograph showing $4$ of the $35$ hybrid structures present on the chip used in the experiment. (b) Magnified view of (a) showing the details of the hybrid structure (false colors). (c) and (d) Thermal displacement spectra $S_{xx}^{1/2}\left[\Omega\right]$  of the fundamental out-of-plane (mode 1) and in-plane (mode 2) modes, respectively. The green curves denote the thermal motion of the mechanical modes. The insets show finite-element simulations of the respective mechanical modes with the red arrows representing the direction of displacement. The gray area in (d) represents the bandwidth $\Delta\nu_{\rm{D}}$ in which the stabilization scheme enables noise cancellation at the fundamental thermodynamical limit. (e) Experimental setup for the stabilization mechanism, with the readout and actuation of the mechanical oscillator being obtained via optical means with external cavity diode lasers (ECDL). The readout signal from the photodiode (PD) is split into a demodulation circuit for the directly driven noise detector mode 2 and a circuit for mode 1, which is either a directly driven (straight line) or self-driven (gray line) circuit. Two amplitude modulators provide the means to drive the mechanical modes (AM2) and to control the mechanical frequency by changing the optical power coupled into the cavity (AM1). The frequency fluctuations $\delta\nu_{\rm{M},1}$ and $\delta\nu_{\rm{M},2}$ are recorded either with an oscilloscope, when the system is operated in the direct driving mode or $\delta\nu_{\rm{M},1}$ is recorded with a frequency counter when the mode is self-driven.}%
\label{Fig1}%
\end{figure}

\clearpage

\begin{figure}[ptb]
\includegraphics[scale=.32]{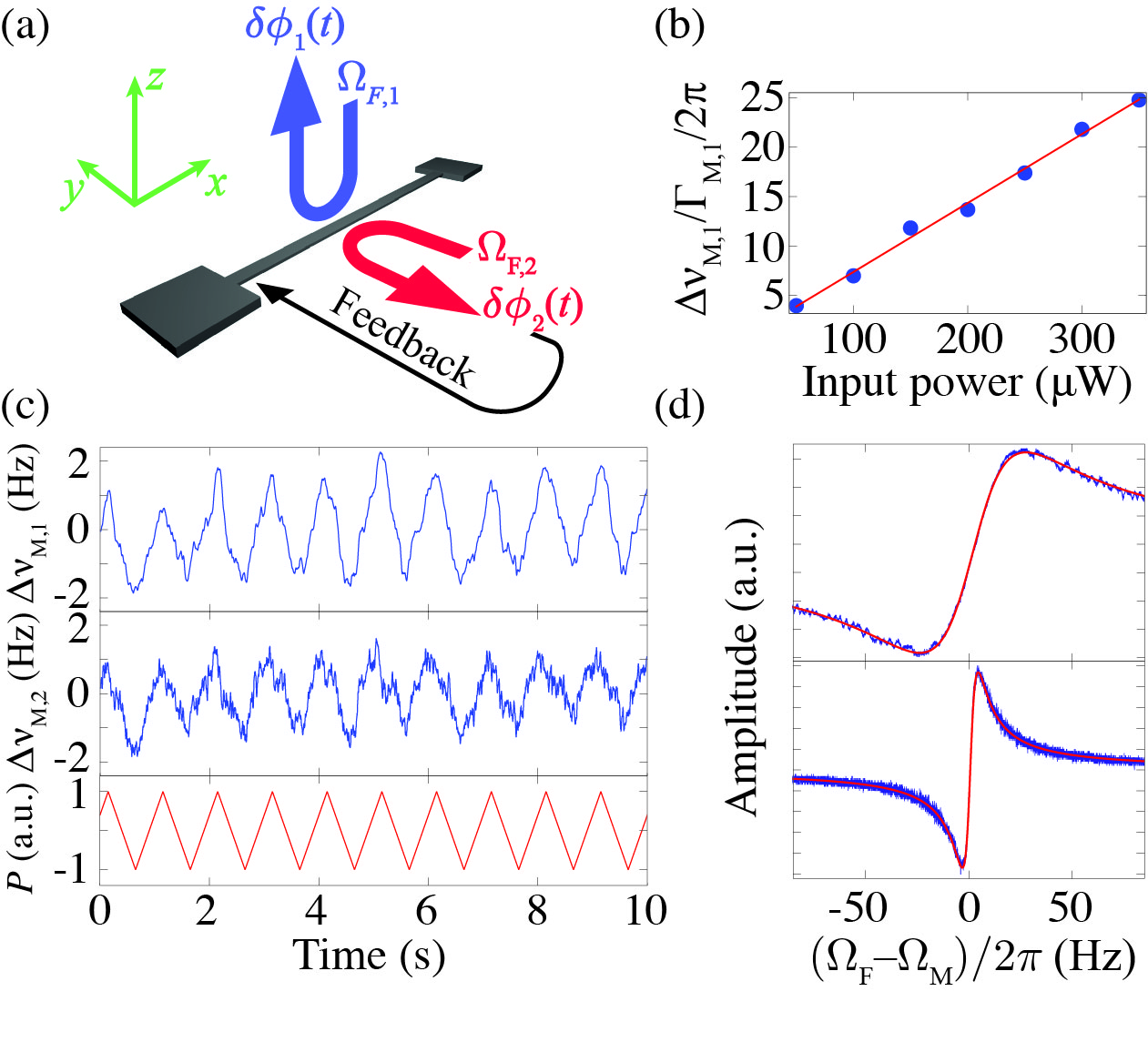}
\caption{\textbf{Detecting and stabilizing frequency noise with an optonanomechanical system.} (a) Scheme describing the stabilization principle. Phase fluctuations of the in-plane mode (mode 2) are detected by determining its out-of-phase response to a coherent drive and serve as an error signal. By applying a feedback control based on these fluctuations, the out-of-plane mode (mode 1) is stabilized. (b) Resonance frequency shift $\Delta\nu_{\rm{M},1}$ of mode 1 in terms of its mechanical linewidth $\Gamma_{\rm{M,1}}$ versus input optical power. (c) Resonance frequency shifts $\Delta\nu_{\rm{M},1}$ and $\Delta\nu_{\rm{M},2}$ of modes 1 and 2, respectively, for a saw-tooth modulation of the input optical power $P$. The correlation of the response of both modes is clearly visible. (d) The out-of-phase mechanical response of mode 2 (upper graph) and mode 1 (lower graph) to a coherent drive versus the detuning $(\Omega_{\mathrm{F}}-\Omega_{\mathrm{M}})/2\pi$.}%
\label{Fig2}%
\end{figure}

\clearpage

\begin{figure}[ptb]
\includegraphics[scale=.2]{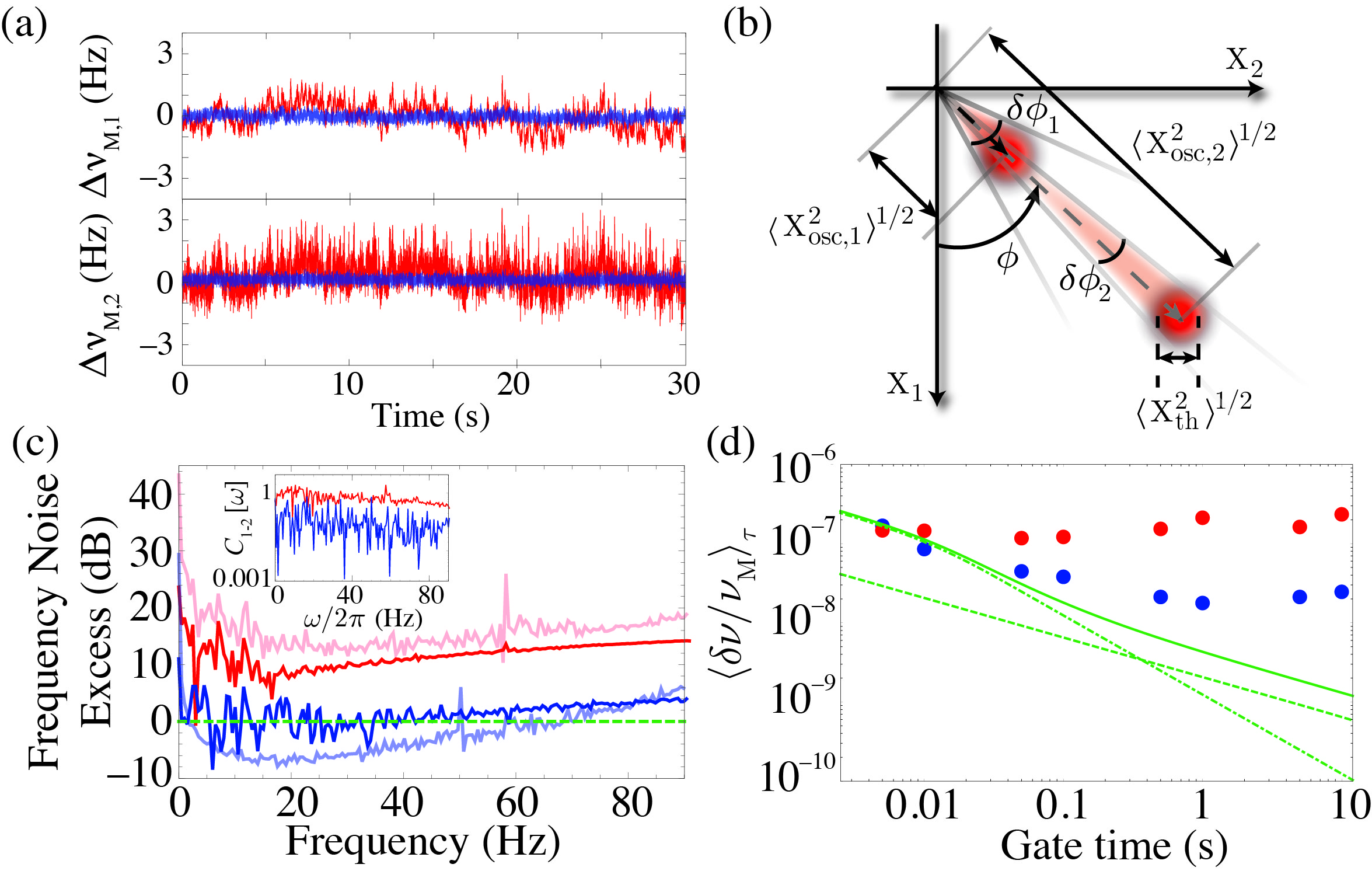}
\caption{\textbf{Frequency stabilization at the ultimate thermodynamic limit.} (a) Time evolution of the frequency drift $\delta\nu_{\rm{M}}$ for mode 1 (upper graph) and mode 2 (lower graph). The non-stabilized case is given by the red curve and the stabilized case by the grey one. (b) Thermal equivalent phase noise in phase-space: $X_1$ and $X_2$ are the in-phase and out-of-phase components of mechanical motion, respectively, and form the phase-space. In this representation, the phase of mechanical motion identifies to the azimuthal coordinate $\phi$. The phase noise $\delta\phi$ simply corresponds to the angle under which the thermal distribution appears from the center. (c) Frequency noise excess of modes 1 (dark-colored) and 2 (light-colored) over the Fourier frequency components. The red and blue curves show the non-stabilized and stabilized case, respectively. The stabilization scheme reduces the frequency noise by over 10 dB up to 90 Hz and allows to stabilize mode 1 to the thermal limit for a wide bandwidth. Inset: The correlation of frequency response for both modes for the non-stabilized (red) and the stabilized (blue) case. (d) The fractional Allan deviation for the non-stabilized (red) and stabilized (blue) case. The dashed line shows the thermal limit associated with mode 1 for a self-driven oscillator. The dot-dashed line shows the thermal limit for mode 2, which was externally driven. The full line shows the thermal limit for mode 1 in presence of feedback (Eq.\ref{eq:lim-total}) by taking into account the reinjected imprecision from mode 2.}%
\label{Fig3}%
\end{figure}

\clearpage

\begin{figure}[ht]
\includegraphics[scale=.18]{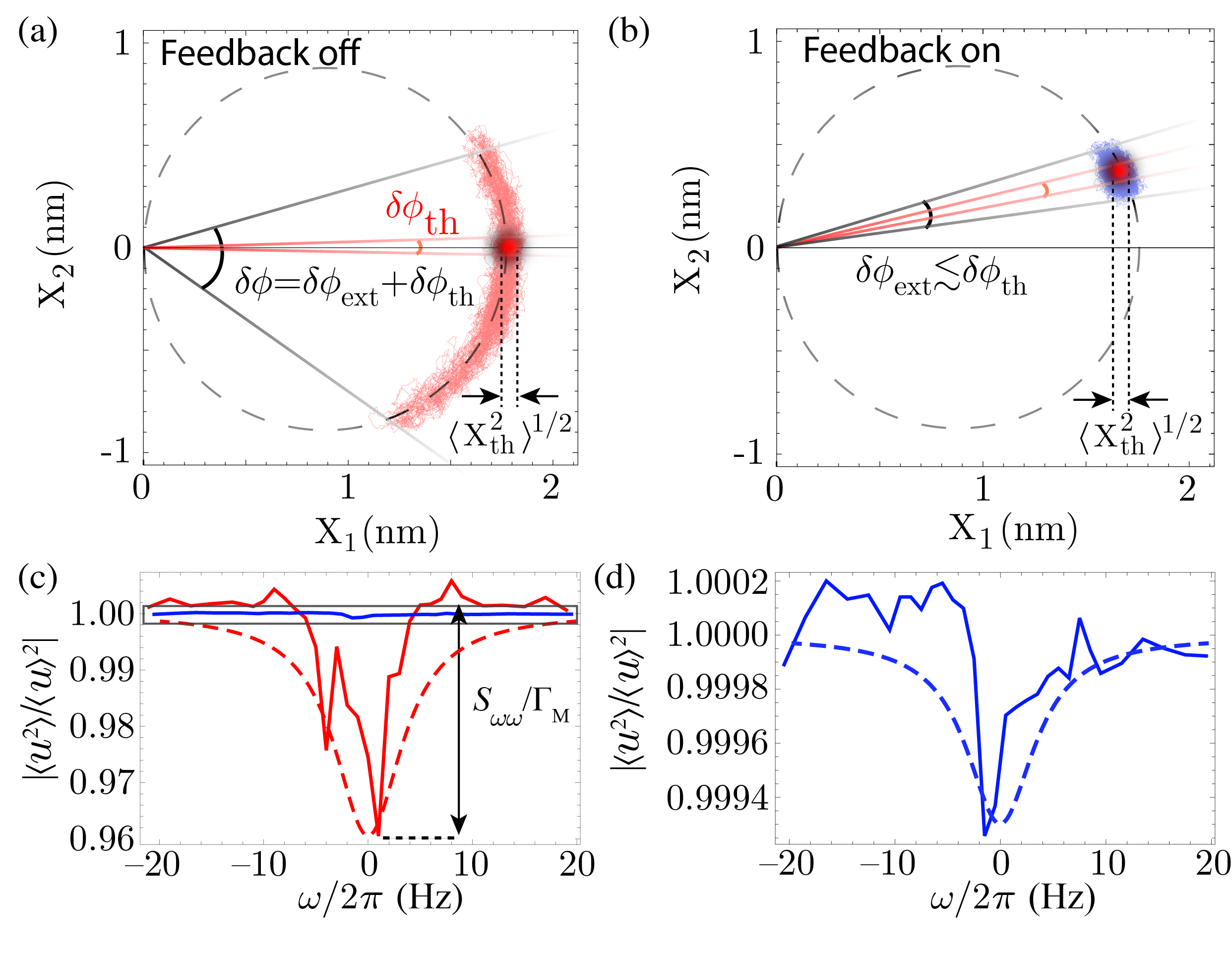}
\caption{\textbf{Frequency noise self-characterization.} (a) A realization of the phase-space trajectory acquired in the non-stabilized case when driving the nanomechanical beam at the mechanical resonance frequency ($\omega\simeq0$). The acquisition duration is on the order of one minute. The red Gaussian disk represents the thermal distribution of variance $\langle X_{\mathrm{th}}^2\rangle$, responsible for an equivalent phase noise $\delta\phi_{\mathrm{th}}$.(b) Same as in (a), but for the stabilized case. The remaining phase noise clearly appears as being substantially reduced compared to non-stabilized case. (c) The ratio $\langle u^{2}\rangle/\langle u\rangle^{2}$ for mode 1, which signifies the presence of frequency fluctuations due to non-thermal contributions, versus  detuning from resonance frequency (red: nonstabilized; blue: stabilized). The red dashed-line corresponds to the model given by Eq. \ref{eq:dykmanrig} for a Gaussian noise with frequency power spectral density $15\,\mathrm{dB}$ above the thermal limit. Stabilization greatly reduces the dip due to excessive non-thermal noise. (d) Detailed view of $\langle u^{2}\rangle/\langle u\rangle^{2}$ for the stabilized case. The blue dasehd-line correspond to the model given by Eq. \ref{eq:dykmanrig} for a Gaussian noise with frequency power spectral density $1.5\,\mathrm{dB}$ above the thermal limit.}%
\label{Fig4}%
\end{figure}

\clearpage
\newpage

\renewcommand{\thefigure}{S\arabic{figure}}
\renewcommand{\thetable}{S\arabic{table}}
\renewcommand{\theequation}{$\mathrm{S\,} $\arabic{equation}}
\setcounter {figure} {0}
\setcounter {equation} {0}
\setcounter {table} {0}
\begin{center}
\large{\textbf{
Supplementary Material - Stabilization of a strongly driven nanomechanical oscillator to the thermal limit}}
\end{center}
\vspace{.2in}

\setcounter{page}{11}

\section{Phase noise spectrum of an oscillator at thermal equilibrium}
In this section, we present the calculation of a thermal equivalent phase noise for a mechanical resonator at thermal equilibrium. Both cases of a free running and externally driven oscillator are treated successively. In the following, we assume the oscillator to be well described within the weakly damped harmonic approximation, with mechanical susceptibility $\chi[\Omega]$ given by:
\begin{eqnarray}
\chi[\Omega]&=&\frac{1}{M\left(\Omega_{\mathrm{M}}^2-\Omega^2-i\Gamma_{\mathrm{M}}\Omega\right)},\label{eq:suscpetibility}
\end{eqnarray}
with $M$, $\Omega_{\mathrm{M}}$ and $\Gamma_{\mathrm{M}}$ as the mass, mechanical resonance frequency and damping rate respectively.
\subsection{Free-running thermally driven oscillator}
\label{section:1-1}

\begin{figure}[ht]
\includegraphics[width=\columnwidth]{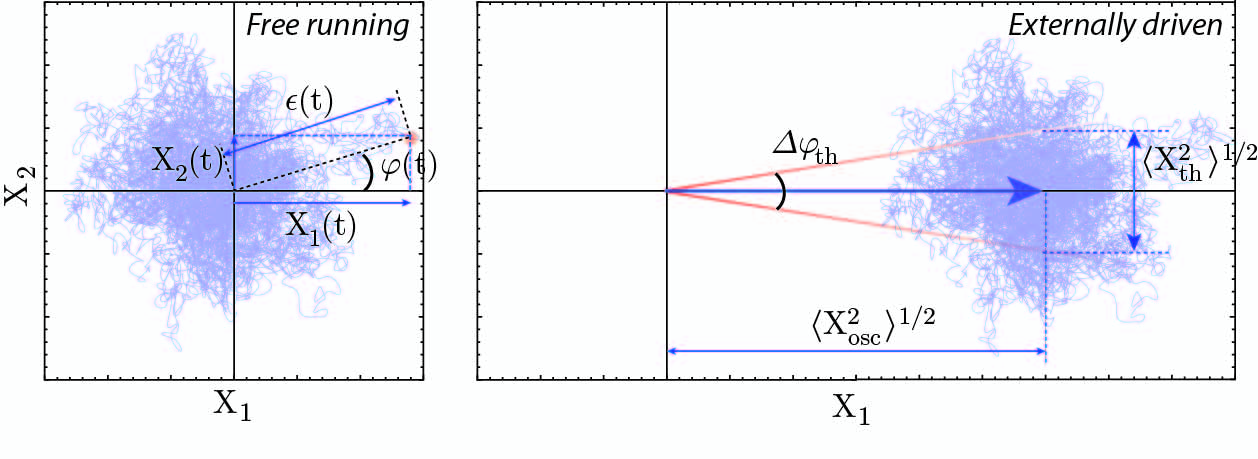} \centering
\caption{Phase-space representation of a mechanical state at thermal equilibrium. (a) Phase-space trajectory of a mechanical oscillator left at thermal equilibrium. The trajectory describes in cartesian coordinates $(X_1(t),X_2(t))$, or equivalently in polar coordinates $(\epsilon(t),\varphi(t))$. (b) Phase-space trajectory of a mechanical resonator driven at the mechanical resonance. The thermal state is "displaced" of a quantity $\langle X_{\mathrm{osc}}^3\rangle^{1/2}$ along the abscise.}%
\label{Fig1}%
\end{figure}

We start this calculation with noting that the phase $\varphi(t)$ of a thermal state $x(t)$ can be expressed as a function of its slowly varying in-phase and out-of-phase motion components $X_1(t)$ and $X_2(t)$ ($x(t)=X_1(t)\cos{\Omega_{\mathrm{M}}t}+X_2(t)\sin{\Omega_{\mathrm{M}}t}$, $\Omega_{\mathrm{M}}/2\pi$ denoting the mechanical resonance frequency) as follows:
\begin{eqnarray}
\varphi(t)&=&2\arctan\left(\frac{X_2(t)}{X_1(t)}\right).\label{eq:2-1-1}
\end{eqnarray}
Here the factor of $2$ preceding the $\arctan$ function is to allow the phase to vary over the interval $[0,2\pi]$. In order to obtain a suitable expression for deriving the spectral properties of $\varphi$, we subsequently take the derivative of Eq.\ref{eq:2-1-1}, which gives:
\begin{eqnarray}
\frac{1}{2}\epsilon(t)\dot{\varphi}(t)&=&\dot{X}_1(t)X_2(t)-\dot{X}_2(t)X_1(t),\label{eq:2-1-2}
\end{eqnarray}
with $\epsilon(t)=X_1^2(t)+X_2^2(t)$ being proportional to the energy operator. The phase and the energy being independent, the autocorrelation of the quantity $\epsilon\dot{\varphi}$ can be factorized, which leads to:
\begin{eqnarray}
\frac{1}{4}\langle\epsilon(t)\epsilon(t')\rangle\langle\dot{\varphi}(t)\dot{\varphi}(t')\rangle&=&\langle\dot{X}_1(t)\dot{X}_1(t')\rangle\langle X_2(t)X_2(t')\rangle+\langle\dot{X}_2(t)\dot{X}_2(t')\rangle\langle X_1(t)X_1(t')\rangle\nonumber\\
&&-\langle\dot{X}_1(t)X_1(t')\rangle\langle X_2(t)\dot{X}_2(t')\rangle-\langle\dot{X}_1(t')X_1(t)\rangle\langle X_2(t')\dot{X}_2(t)\rangle.\label{eq:2-1-3}
\end{eqnarray}
In the above equation, $\langle...\rangle$ is for statistical average, $t$ and $t'$ for two arbitrary moments in time, and the factorization on the right side arises from the well known statistical independence of the motion quadratures $X_1$ and $X_2$. For evident reasons of symmetry, the two terms on the right side of Eq. \ref{eq:2-1-3} on the first line (resp. on the second line) are identical such that it can be written:
\begin{eqnarray}
\frac{1}{8}\langle\epsilon(t)\epsilon(t')\rangle\langle\dot{\varphi}(t)\dot{\varphi}(t')\rangle&=&\langle\dot{X}_1(t)\dot{X}_1(t')\rangle\langle X_1(t)X_1(t')\rangle-\langle\dot{X}_1(t)X_1(t')\rangle\langle X_1(t)\dot{X}_1(t')\rangle.\label{eq:2-1-4}
\end{eqnarray}
 We now determine each factor appearing on the ride side of Eq. \ref{eq:2-1-4}. We start with the simpler one $\langle X_1(t)X_1(t')\rangle$, which identifies with the autocorrelation function of the quadrature $X_1$. Assuming the stationarity of the problem, we have $\langle X_1(t)X_1(t')\rangle=\langle X_1(0)X_1(\tau=t'-t)\rangle$, and the analytic expression of this autocorrelation function can be obtained via the Fourier transform of the spectrum of $X_1$ $S_{X,1}[\omega]=|\chi[\Omega_{\mathrm{M}}+\omega]|^2S_{\mathrm{F}}^{\mathrm{T}}[\omega]$ (where $S_{\mathrm{F}}^{\mathrm{T}}[\omega]=4 M\Gamma_{\mathrm{M}} k_B T$ is the thermal force spectral density):
\begin{eqnarray}
\langle X_1(0)X_1(\tau)\rangle&=&\int_{-\infty}^{+\infty}\mathrm{d}\omega e^{i\omega\tau}S_{X,1}[\omega]\nonumber\\
&=&\frac{S_{\mathrm{F}}^{\mathrm{T}}}{(2M\Omega_{\mathrm{M}})^2\Gamma_{\mathrm{M}}}e^{-\frac{\Gamma_{\mathrm{M}}}{2}|t|}.\label{eq:2-1-5}
\end{eqnarray}
Similarly, noting that $S_{\dot{X},1}[\omega]=\omega^2S_{X,1}[\omega]$, one obtains the autocorrelation function of $\dot{X}_1$:
\begin{eqnarray}
\langle\dot{X}_1(0)\dot{X}_1(\tau)\rangle&=&\frac{S_{\mathrm{F}}^{\mathrm{T}}}{(2M\Omega_{\mathrm{M}})^2}\left(\delta(t)-\frac{\Gamma_{\mathrm{M}}}{4}e^{-\frac{\Gamma_{\mathrm{M}}}{2}|t|}\right),\label{eq:2-1-6}
\end{eqnarray}
where $\delta$ is for the Dirac delta function.
To determine the cross correlations $\langle \dot{X}_1X_1\rangle$ and $\langle X_1\dot{X}_1\rangle$, we write the quadrature $X_1$ as the convolution between its impulse response $\chi(t)=\int_{-\infty}^{+\infty}\mathrm{d}\omega e^{i\omega t}\chi[\Omega_{\mathrm{M}}+\omega]=1/(2M\Omega_{\mathrm{M}})\Theta(t)e^{-\frac{\Gamma_{\mathrm{M}}}{2}t}$ ($\Theta$ denoting the Heaviside step function) and an effective thermal force $F_{\mathrm{th}}$ (with $\langle F_{\mathrm{th}}(t)F_{\mathrm{th}}(t')\rangle= S_{\mathrm{F}}^{\mathrm{T}}\delta(t-t')$) \cite{Briant2003}. After a tedious but simple calculation, we obtain:
\begin{eqnarray}
\langle\dot{X}_1(t)X_1(t')\rangle\langle X_1(t)\dot{X}_1(t')\rangle&=&-\frac{(S_{\mathrm{F}}^{\mathrm{T}})^2}{(2M\Omega_{\mathrm{M}})^2}\times\frac{e^{-\Gamma_{\mathrm{M}}|t|}}{4}.\label{eq:2-1-7}
\end{eqnarray}
Combining Eqs. \ref{eq:2-1-4}, \ref{eq:2-1-5}, \ref{eq:2-1-6}, and \ref{eq:2-1-7}, we finally find:
\begin{eqnarray}
\frac{1}{8}\langle\epsilon(t)\epsilon(t')\rangle\langle\dot{\varphi}(t)\dot{\varphi}(t')\rangle&=&\frac{(S_{\mathrm{F}}^{\mathrm{T}})^2}{(2M\Omega_{\mathrm{M}})^2}
\times\delta(t)e^{-\frac{\Gamma_{\mathrm{M}}}{2}|t|}\label{eq:2-1-8}.
\end{eqnarray}
The autocorrelation of $\epsilon$ remains to be calculated. To do so, we first note that $\langle\epsilon(0)\epsilon(\tau)\rangle=2\left(\langle X_1^2(0)X_1^2(\tau)\rangle+\langle X_1^2(0)\rangle^2\right)$. The equipartition of the energy gives immediately the second term of the previous parenthesis, $\langle X_1^2(0)\rangle^2=(k_BT/M\Omega_{\mathrm{M}}^2)^2$. The first one, that is the autocorrelation of $X_1^2$, remains to be determined, which we do starting with expanding it as:
\begin{eqnarray}
\langle X_1^2(0)X_1^2(\tau)\rangle&=&\iiiint\limits_{[-\infty,+\infty]^4}\mathrm{d}t_1\mathrm{d}t_2\mathrm{d}t_3\mathrm{d}t_4\chi(0-t_1)\chi(0-t_2)\chi(\tau-t_3)\chi(\tau-t_4)\langle F_{\rm{th}}(t_1)F_{\rm{th}}(t_2)F_{\rm{th}}(t_3)F_{\rm{th}}(t_4)\rangle,\nonumber\\
\label{eq:moment2}
\end{eqnarray}
$\left(F_{\rm{th}}(t_1),F_{\rm{th}}(t_2),F_{\rm{th}}(t_3),F_{\rm{th}}(t_4)\right)$ being a multivariate Gaussian distribution, Wick's theorem applies, and the statistical average of Eq.\ref{eq:moment2} can be rewritten as:
\begin{eqnarray}
\langle F_{\rm{th}}(t_1)F_{\rm{th}}(t_2)F_{\rm{th}}(t_3)F_{\rm{th}}(t_4)\rangle&=&\langle F_{\rm{th}}(t_1)F_{\rm{th}}(t_2)\rangle\langle F_{\rm{th}}(t_3)F_{\rm{th}}(t_4)\rangle+\langle F_{\rm{th}}(t_1)F_{\rm{th}}(t_3)\rangle\langle F_{\rm{th}}(t_2)F_{\rm{th}}(t_4)\rangle\nonumber\\
&&+\langle F_{\rm{th}}(t_1)F_{\rm{th}}(t_4)\rangle\langle F_{\rm{th}}(t_2)F_{\rm{th}}(t_3)\rangle.\label{eq:corr4}
\end{eqnarray}
The previous equation shows that the fourth moment of the thermal force is a combination of products of its autocorrelation functions $\langle F_{\rm{th}}(t_i)F_{\rm{th}}(t_j)\rangle=S_{\mathrm{F}}^{\mathrm{T}}[\Omega\simeq\Omega_{\mathrm{m}}]\delta(t_i-t_j)$:
\begin{eqnarray}
\langle F_{\rm{th}}(t_1)F_{\rm{th}}(t_2)F_{\rm{th}}(t_3)F_{\rm{th}}(t_4)\rangle&\simeq&S_{\mathrm{FF}}^{\mathrm{th}}[\Omega_{\mathrm{m}}]^2\left\{\delta(t_1-t_2)\delta(t_3-t_4)+\delta(t_1-t_3)\delta(t_2-t_4)\right.\nonumber\\
&&\,\,\,\,\,\,\,\,\,\,\,\,\,\,\,\,\,\,\,\,\,\,\,\,\left.+\delta(t_1-t_4)\delta(t_2-t_3)\right\}.\label{eq:corr4-bis}
\end{eqnarray}
Replacing Eq.\ref{eq:corr4-bis} in Eq.\ref{eq:moment2}, one obtains:
\begin{eqnarray}
\langle X_1^2(0)X_1^2(\tau)\rangle
&=&(S_{\mathrm{F}}^{\mathrm{T}})^2\Bigg\{\left(\int_{-\infty}^{+\infty}\mathrm{d}t_1\chi^2(t_1)\right)^2+2 \left(\int_{-\infty}^{+\infty}\mathrm{d}t_1\chi(t_1)\chi(\tau-t_1)\right)^2\Bigg\}.\label{eq:corr4-ter}
\end{eqnarray}
Using the expression of the impulse response $\chi(t)$ given above, one finally obtains for the autocorrelation of $\epsilon$:
\begin{eqnarray}
\langle\epsilon(0)\epsilon(\tau)\rangle&=&4\left(\frac{k_BT}{M\Omega_{\mathrm{M}}^2}\right)^2\left(1+e^{-\Gamma_{\mathrm{M}}|t|}\right).\label{eq:2-1-9}
\end{eqnarray}
Last, using that $S_{\dot{\varphi}\dot{\varphi}}[\omega]=\omega^2S_{\varphi\varphi}[\omega]$, Eqs. \ref{eq:2-1-8} and \ref{eq:2-1-9} lead to the very simple expression for the phase noise:
\begin{eqnarray}
S_{\varphi\varphi}[\omega]&=&\frac{\Gamma_{\mathrm{M}}}{\omega^2}.\label{eq:2-1-10}
\end{eqnarray}
The phase noise of a free-running, thermally driven mechanical resonator is hence entirely determined by and proportional to its dissipation rate, which is a well known result. Note however that importantly, this does not depend on the temperature.
\subsection{Feedback driven oscillator at thermal equilibrium}
\label{section:1-2}
It is straight to show that the expression \ref{eq:2-1-10} extends to the case of a feedback driven oscillator \emph{at thermal equilibrium}: Indeed, the effect of linear dissipative feedback is to change the susceptibility to an effective susceptibility $\chi_{\mathrm{fb}}[\omega]$ with the linewidth $\Gamma_{\mathrm{M}}$ being replaced by an effective linewidth $\Gamma_{\mathrm{fb}}$ which is an affine function of feedback gain. Noting that the thermal force is unchanged by feedback, and assuming that the feedback noise is negligible, the phase noise of the feedback driven resonator is obtained by deriving the above analysis with replacing $\Gamma_{\mathrm{M}}$ by $\Gamma_{\mathrm{fb}}$, which leads to:
\begin{eqnarray}
S_{\varphi\varphi}^{\mathrm{fb}}[\omega]&=&\frac{\Gamma_{\mathrm{fb}}}{\omega^2}.\label{eq:2-1-11}
\end{eqnarray}
Thereby, positive dissipative strong feedback ($\Gamma_{\mathrm{eff}}\ll\Gamma_{\mathrm{M}}$) enables considerable reduction of the phase noise of the resonator. While this result is somehow similar to the situation of an externally driven resonator as we will see in the next section, the phase noise decrease is limited to the stability of the external reference, while feedback is self-referenced. The experimental performance of the feedback technique will therefore mostly rely on the ability to design a low-noise feedback circuit.
\subsection{Externally driven resonator}
\label{section:1-3}
To do so, we start back with Eq. \ref{eq:2-1-1}

We now turn to the case of an externally driven mechanical resonator. We further assume that the oscillator is resonantly driven using a coherent amplitude $\langle X_{\mathrm{osc}}^2\rangle^{1/2}\gg(k_B T/M\Omega_{\mathrm{M}}^2)^{1/2}$. As a consequence, since $X_1$ has been arbitrarily defined as being the phase reference (see Fig. \ref{Fig1}), its (random) thermal component can be neglected $X_{\mathrm{th},1}$ ($X_1(t)=\langle X_{\mathrm{osc}}^2\rangle^{1/2}+X_{\mathrm{th},1}(t)\simeq\langle X_{\mathrm{osc}}^2\rangle^{1/2}$). The motion phase imprecision resulting from the thermal fluctuations $X_{\mathrm{th},2}$ of the quadrature $X_2$ is thereby given by:
\begin{eqnarray}
\varphi(t)\simeq\frac{X_{\mathrm{th},2}(t)}{\langle X_{\mathrm{osc}}^2\rangle^{1/2}},\label{eq:2-1-12}
\end{eqnarray}
The phase noise associated to an externally driven state at thermal equilibrium therefore follows immediately from Eq. \ref{eq:2-1-12}:
\begin{eqnarray}
S_{\varphi\varphi}^{\mathrm{osc}}[\omega]&=&\frac{S_{\mathrm{XX}}^{\mathrm{th}}[\omega]}{\langle X_{\mathrm{osc}}^2\rangle},\label{eq:2-1-13}
\end{eqnarray}
which is a well known result \cite{Albrecht1991,Cleland2002}: As increasing the external drive amplitude\footnote{We assume the mechanical motion to remain linear.}, the thermal induced phase noise decreases. However, as noted above, the frequency stability limit then relies on that of the external reference. Reaching better stability performance will be thereby preferably possible using the positive feedback technique as announced above.
\begin{figure}[ht]
\includegraphics[width=\columnwidth]{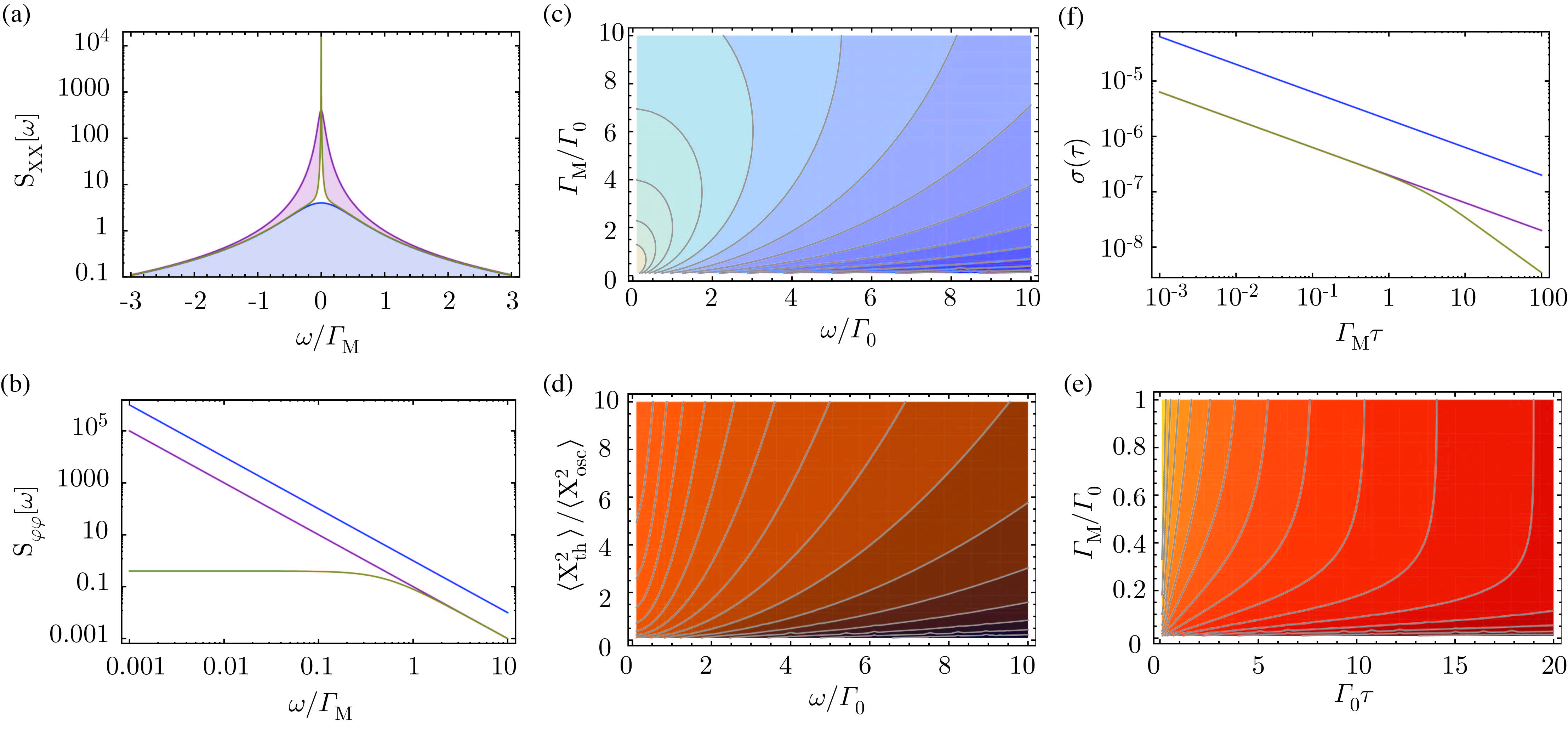} \centering
\caption{Phase and frequency stability of a mechanical resonator at thermodynamic equilibrium. (a) Schematic of the displacement spectrum of a thermally-driven oscillator (blue), feedback driven resonator with $\langle X_{\mathrm{fb}}^2\rangle=\langle X_{\mathrm{th}}^2\rangle$, and externally driven resonator (brown) with $\langle X_{\mathrm{osc}}^2\rangle=\langle X_{\mathrm{th}}^2\rangle$. (b) Corresponding phase noise spectra. (c) Phase noise of an externally driven high-$Q$ mechanical oscillator as a function of the Fourier frequency and mechanical damping-rate (normalized to an arbitrary damping unit $\Gamma_0$). (d) Phase noise of an externally driven high-$Q$ mechanical oscillator as a function of the Fourier frequency and driving power. In both (c) and (d), the phase noise decreases towards darker regions. (c) and (d) show that increasing the external drive or mechanical quality factor lead to similar improvement for the phase noise performance. (e) Allan deviation of a thermally-driven oscillator (blue), feedback driven resonator with $\langle X_{\mathrm{fb}}^2\rangle=\langle X_{\mathrm{th}}^2\rangle$, and externally driven resonator (brown) with $\langle X_{\mathrm{osc}}^2\rangle=\langle X_{\mathrm{th}}^2\rangle$. (f) Allan deviation for an externally driven resonator as a function of gate time and mechanical damping rate (normalized to an arbitrary damping unit $\Gamma_0$). The Allan deviation decreases towards the darker regions. Note the contours which tend to be parallel at higher damping rates, which reflects the fact that the frequency stability becomes independent of damping rate $\Gamma_{\mathrm{M}}$ at higher gate times.}%
\label{Fig2}%
\end{figure}
\section{Thermodynamic limit of frequency stability}
In this section, we first establish the mathematical connection between phase noise and frequency stability. We then derive the analytical expression of the Allan deviation corresponding to the three cases treated above.
\subsection{Allan variance as a function of phase noise}
Here we establish the mathematical relationship between the Allan variance and the phase noise. For a given gate time $\tau$, the Allan variance $\sigma(\tau)$ defines as:
\begin{eqnarray}
\sigma^2(\tau)&=&\frac{1}{2f_{\rm M}^2}\lim_{n\rightarrow\infty}\left\{\frac{1}{n-1}\sum_{k=2}^n\Big(\bar{f}_k-\bar{f}_{k-1}\Big)^2\right\}\nonumber\\
&=&\frac{1}{2f_{\rm M}^2}\lim_{n\rightarrow\infty}\left\{\frac{1}{n-1}\sum_{k=2}^n\Bigg(\frac{1}{\tau}\int\limits_{k\tau}^{(k+1)\tau}\mathrm{d} t f(t)-\frac{1}{\tau}\int\limits_{(k-1)\tau}^{k\tau}\mathrm{d} t f(t)\Bigg)^2\right\},
\end{eqnarray}
where $f_{\mathrm{M}}=\Omega_{\mathrm{M}}/2\pi$ denotes the resonator's mechanical resonance frequency. expanding the squared parenthesis in the above equation, one obtains:
\begin{eqnarray}
\sigma^2(\tau)&=&\frac{1}{2f_{\rm M}^2\tau^2}\lim_{n\rightarrow\infty}\left\{\frac{1}{n-1}\sum_{k=2}^n\Bigg(\,\,
\iint\limits_{\mathcal{D}_{k+1,k+1}}\mathrm{d}t\mathrm{d}t'f(t)f(t')+
\iint\limits_{\mathcal{D}_{k,k}}\mathrm{d}t\mathrm{d}t'f(t)f(t')-
2\iint\limits_{\mathcal{D}_{k+1,k}}\mathrm{d}t\mathrm{d}t'f(t)f(t')\Bigg)\right\},\nonumber\\
\end{eqnarray}
where $\mathcal{D}_{j_1,j_2}=[(j_1-1)\tau,j_1\tau]\times[(j_2-1),j_2\tau]$. The latter equation can be rewritten using fixed integration's boundaries as follows:
\begin{eqnarray}
\sigma^2(\tau)&=&\frac{1}{2f_{\rm M}^2\tau^2}\lim_{n\rightarrow\infty}\left\{\frac{1}{n-1}\sum_{k=2}^n
\Bigg(\,\,\iint\limits_{[0,\tau]^2}\mathrm{d}t\mathrm{d}t'f(t+k\tau)f(t'+k\tau)+\iint\limits_{[0,\tau]^2}\mathrm{d}t\mathrm{d}t'f(t+(k-1)\tau)f(t'+(k-1)\tau)\right.\nonumber\\
&&\left.\,\,\,\,\,\,\,\,\,\,\,\,\,\,\,\,\,\,\,\,\,\,\,\,\,\,\,\,\,\,\,\,\,\,\,\,\,\,\,\,\,\,\,\,\,\,\,\,\,\,\,\,
\,\,\,\,\,\,\,\,\,\,-2\iint\limits_{[0,\tau]^2}\mathrm{d}t\mathrm{d}t'f(t+k\tau)f(t'+(k-1)\tau)\Bigg)\right\}.
\end{eqnarray}
Swapping the statistical and temporal summations and using the ergodic assumption, one obtains:
\begin{eqnarray}
\sigma^2(\tau)&=&\frac{1}{f_{\rm M}^2\tau^2}\iint\limits_{[0,\tau]^2}\mathrm{d}t\mathrm{d}t'\Big(\langle f(t)f(t')\rangle-\langle f(t)f(t'-\tau)\rangle\Big).
\end{eqnarray}
To express the Allan variance in terms of frequency noise $S_{\mathrm{ff}}[\Omega]$, we use the Wiener-Khintchin theorem which enables expressing the autocorrelation functions featured in the above expression as the inverse Fourier Transform of the spectrum, $\langle f(t)f(t')\rangle=\frac{1}{2\pi}\int\limits_{-\infty}^{+\infty}\mathrm{d}\Omega S_{\mathrm{ff}}[\Omega]e^{i\Omega(t-t')}$. The latter equation reads therefore:
\begin{eqnarray}
\sigma^2(\tau)&=&\frac{1}{2\pi f_{\rm M}^2\tau^2}\iint\limits_{[0,\tau]^2}\mathrm{d}t\mathrm{d}t'\int\limits_{-\infty}^{+\infty}\mathrm{d}\Omega S_{\mathrm{ff}}[\Omega]e^{i\Omega(t'-t)}\Big(1-e^{-i\Omega\tau}\Big).
\end{eqnarray}
Fubini's theorem enables swapping the time and frequency integration to finally have:
\begin{eqnarray}
\sigma^2(\tau)&=&\frac{1}{\pi f_{\rm M}^2\tau^2}\int\limits_{-\infty}^{+\infty}\mathrm{d}\Omega\frac{S_{\mathrm{ff}}[\Omega]}{\Omega^2}\Big(1-e^{-i\Omega\tau}\Big)\Big(1-\cos(\Omega\tau)\Big).
\end{eqnarray}
Using that the imaginary part of the latter equation cancels, that the frequency spectrum is a pair function $S_{\mathrm{ff}}[\Omega]=S_{\mathrm{ff}}[-\Omega]$, and that the frequency noise $S_{\phi\phi}[\Omega]=S_{\mathrm{ff}}[\Omega]/\Omega^2$, the Allan variance finally reads:
\begin{eqnarray}
\sigma^2(\tau)&=&\frac{2}{\pi}\left(\frac{2}{\Omega_{\mathrm{M}}\tau}\right)^2\int\limits_0^{+\infty}\mathrm{d}\Omega S_{\varphi\varphi}[\Omega]\sin^4\Big(\frac{\Omega\tau}{2}\Big).\label{eq:general}
\end{eqnarray}
Though being similar to what can be found in Cleland \emph{et al} \cite{cleland2002noise,cleland2005thermomechanical}, this expression is a factor of $\pi$ lower, which is of great importance: as an example, when the first one leads to the conclusion that a system is at the thermal limit, we find that it is actually still a factor $>3$ above.

\subsection{Frequency stability of a mechanical resonator at thermodynamic equilibrium}
Using the result given by Eq. \ref{eq:general}, we are now able to determine the frequency stability associated to the cases treated in sections \ref{section:1-1}, \ref{section:1-2} and section \ref{section:1-3}.
\paragraph{Free-running thermally driven oscillator}
Using Eqs. \ref{eq:2-1-10} and \ref{eq:general}, one obtains for the frequency Allan deviation:
\begin{eqnarray}
\sigma^2(\tau)&=&\frac{1}{Q_{\mathrm{M}}\Omega_{\mathrm{M}}\tau},\label{eq:stab-free-running}
\end{eqnarray}
with $Q_{\mathrm{M}}=\Omega_{\mathrm{M}}/\Gamma_{\mathrm{M}}$ denoting the mechanical quality factor of the resonator. Eq. \ref{eq:stab-free-running} shows that the stability increases with mechanical quality factor, gate time, and frequency: Therefore, besides being of high interest for purposes such as mass sensing \cite{Feng2008,Chaste2012}, high-frequency, high-Q mechanical resonators are extremely useful for absolute stability applications such as time and frequency control \cite{Nguyen2007}.
\paragraph{Feedback driven oscillator at thermal equilibrium}
As suggested by the very similar mathematical expression of the phase noise in the free-running and feedback driven configurations, the expression of the thermal limited frequency stability in the latter case takes the same form as Eq. \ref{eq:stab-free-running}:
\begin{eqnarray}
\sigma^2(\tau)&=&\frac{1}{Q_{\mathrm{fb}}\Omega_{\mathrm{M}}\tau},\label{eq:stab-fb-1}
\end{eqnarray}
with $Q_{\mathrm{fb}}=\Omega_{\mathrm{M}}/\Gamma_{\mathrm{fb}}$ the effective mechanical quality factor\footnote{We assume the feedback mechanism to be purely dissipative, such that it does not induce any change in the mechanical resonance frequency.}. We note that Eq. \ref{eq:stab-fb-1} can be expressed differently, as a function of the feedback driven motion variance $\langle X_{\mathrm{th,fb}}^2\rangle$:
\begin{eqnarray}
\sigma_{\mathrm{fb}}^2(\tau)&=&\frac{\langle X_{\mathrm{th}}^2\rangle}{\langle X_{\mathrm{th,fb}}^2\rangle}\times\frac{1}{Q_{\mathrm{M}}\Omega_{\mathrm{M}}\tau},\label{eq:stab-fb-2}
\end{eqnarray}
where we used that $\Gamma_{\mathrm{fb}}/\Gamma_{\mathrm{M}}=\langle X_{\mathrm{th}}^2\rangle/\langle X_{\mathrm{th,fb}}^2\rangle$ \cite{Gavartin2012}. This expression identifies to the one we derived in the next paragraph in the case of an externally driven oscillator at short gate times ($\tau\leq1/\Gamma_{\mathrm{M}}$), with the important difference that it applies at all gate times (in the limit $\tau\gg1/\Omega_{\mathrm{M}}$). Again, the important difference with external driving is that this stability is no longer limited to the one of the external reference, which is found to be routinely on the order of a few $ppb\,(10^{-9})$ at second-scale for quartz oscillators. Reaching higher stability that become comparable to atomic clocks' (i.e. below $10^{-10}$) will therefore rather require the use of positive dissipative feedback.
\paragraph{Externally driven resonator}
In the limit of a coherent drive $\langle X_{\mathrm{osc}}^2\rangle\gg\langle X_{\mathrm{th}}^2\rangle$, frequency stability of an externally driven resonator at thermal equilibrium is obtained using Eqs. \ref{eq:2-1-13} and \ref{eq:general}, with $S_{\mathrm{XX}}^{\mathrm{th}}[\omega]$ being given by \cite{briant2003caracterisation}:
\begin{eqnarray}
S_{\mathrm{XX}}^{\mathrm{th}}[\omega]&\simeq&|\chi[\Omega_{\mathrm{M}}+\omega]|^2S_{\mathrm{F}}^{\mathrm{T}},
\end{eqnarray}
We note that in MEMS-NEMS literature, this calculation is mostly achieved using a truncated expression of the phase noise, with the slowest noise component being neglected ($S_{\mathrm{XX}}^{\mathrm{th}}[\omega]\simeq S_{\mathrm{XX}}^{\mathrm{th}}[\omega\gg\Gamma_{\mathrm{M}}]$). This is probably a technical legacy: MEMS technology first provided high-Q low-frequency oscillators, with very long mechanical coherence times $1/\Gamma_{\mathrm{M}}$, easily on the order of a few tens of seconds, along which technical limitations (e.g. thermal drifts, pressure fluctuations etc...) are the prominent sources of frequency noise. However, the recent development of high-frequency NEMS has led to low-noise, fast decay rate oscillators, the former truncation being henceforth obsolete. We thereby compute the integral in Eq. \ref{eq:general} without further assumption, and find:
\begin{eqnarray}
\sigma^2(\tau)&=&\frac{k_BT}{M\Omega_{\mathrm{M}}^2\langle X_{\mathrm{osc}}^2\rangle}\times\frac{3+\exp(-\Gamma_{\mathrm{M}}\tau)-4\exp(-\Gamma_{\mathrm{M}}\tau/2)}{(\Omega_{\mathrm{M}}\tau)^2}\nonumber\\
&=&\frac{\langle X_{\mathrm{th}}^2\rangle}{\langle X_{\mathrm{osc}}^2\rangle}\times H(t).\label{eq:stability-thermal}
\end{eqnarray}

Eq.\ref{eq:stability-thermal} shows that the apparent frequency stability depends of the inverse ratio between the coherent drive and the thermal variance, along the thermal equivalent phase noise dependance. The function $H(t)$ describes the gate-time dependance, and features essentially 2 regimes. For short gate-times ($\tau\ll1/\Gamma_{\mathrm{M}}$), we have $H(\tau)\simeq1/Q_{\mathrm{M}}\Omega_{\mathrm{M}}\tau$, and the Allan variance identifies to the result obtained when neglecting the noise's slow components. In this case, one sees that a high $Q_{\mathrm{M}}$ is favorable to lower measurement imprecision, along the usual optimization discussions. As already mentioned above, we also note that this asymptotic expression is identical to the effective frequency stability previously determined in the case of the feedback driven resonator, with $\langle X_{\mathrm{osc}}^2\rangle$ being replaced by $\langle X_{\mathrm{fb}}^2\rangle$, the use of feedback being though preferable as explained before.

On the other hand, for long gate times, Eq.\ref{eq:stability-thermal} leads to $H(t)\simeq3/\Omega_{\mathrm{M}}^2\tau^2$: The thermally-limited Allan variance no longer depends on the mechanical quality factor, and decreases with the square power of $\tau$ and with the fourth power of the mechanical resonance frequency. This $1/\tau^2$ convergence behavior can be understood since averaging over gate-time greater than $\tau\gg1/\Gamma_{\mathrm{M}}$ means that two consecutive averaged realizations of the frequency measurement are independent due to the random nature of thermal noise, in contrast to the short gate-time case where these two measurements are related via the mechanical coherence time. This result also pleads for developing high frequency oscillators \cite{Chaste2012,Feng2008} as highly accurate frequency references, even disregarding their mechanical quality factor, which should be kept high enough only for readout constraints\footnote{The mechanical quality factor has to be such that the thermal noise can be resolved, since the frequency measurement imprecision will be due to detection background otherwise}.

\end{document}